# Applying Physical-layer Network Coding in Wireless Networks

Shengli Zhang, Soung Chang Liew

*Abstract*—A main distinguishing feature of a wireless network compared with a wired network is its broadcast nature, in which the signal transmitted by a node may reach several other nodes, and a node may receive signals from several other nodes, simultaneously. Rather than a blessing, this feature is treated more as an interference-inducing nuisance in most wireless networks today (e.g., IEEE 802.11). This paper shows that the concept of network coding can be applied at the physical layer to turn the broadcast property into a capacity-boosting advantage in wireless ad hoc networks. Specifically, we propose a physical-layer network coding (PNC) scheme to coordinate transmissions among nodes. In contrast to "straightforward" network coding which performs coding arithmetic on digital bit streams after they have been received, PNC makes use of the additive nature of simultaneously arriving electromagnetic (EM) waves for equivalent coding operation. And in doing so, PNC can potentially achieve 100% and 50% throughput increases compared with traditional transmission and straightforward network coding, respectively, in 1-D regular linear networks with multiple random flows. The throughput improvements are even larger in 2-D regular networks: 200% and 100%, respectively.

*Index Terms* — Network coding, wireless networks, throughput

## I. INTRODUCTION

ONE of the biggest challenges in wireless communication is how to deal with the interference at the receiver when signals from multiple sources arrive simultaneously. In the radio channel of the physical layer of wireless networks, data are transmitted through electromagnetic (EM) waves in a broadcast manner. The interference between these EM waves causes the data to be scrambled.

To overcome its negative impact, most schemes attempt to find ways to either reduce or avoid interference through receiver design or transmission scheduling [1]. For example, in 802.11 networks, the carrier-sensing mechanism allows at most one source to transmit or receive at any time within a carrier-sensing range. This is obviously inefficient when multiple nodes have data to transmit.

While interference causes throughput degradation on wireless networks in general, its negative effect for multi-hop ad hoc networks is particularly significant. For example, in 802.11 networks, the theoretical throughput of a multi-hop flow in a linear network is less than 1/4 of the single-hop case due to the "self interference" effect, in which packets of the same flow but at different hops collide with each other [2, 3].

Instead of treating interference as a nuisance to be avoided, we can actually embrace interference to improve throughput performance with the "right mechanism". To do so in a multi-hop network, the following goals must be met:

1. A relay node must be able to convert simultaneously received signals into interpretable output signals to be relayed to their final destinations.
2. A destination must be able to extract the information addressed to it from the relayed signals.

The capability of network coding to combine and extract information through simple Galois field GF($2^n$) additions [4, 5] provides a potential approach to meet such goals. However, network coding arithmetic is generally only applied on bits that have already been correctly received. That is, when the EM waves from multiple sources overlap and mutually interfere, network coding cannot be used to resolve the data at the receiver. So, criterion 1 above cannot be met.

This paper proposes the application of network coding directly within the radio channel at the physical layer. We call this scheme Physical-layer Network Coding (PNC). The main idea of PNC is to create an apparatus similar to that of network coding, but at the physical layer that deals with EM signal reception and modulation. Through a proper modulation-and-demodulation technique at the relay nodes, additions of EM signals can be mapped to GF($2^n$) additions of digital bit streams, so that the interference becomes part of the arithmetic operation in network coding. The basic idea of PNC was first put forth in our conference paper in [6]. Going beyond [6], this paper addresses a number of practical issues of applying PNC in wireless networks. In particular, we propose specific scheduling algorithms for 1-D and 2-D regular networks that make use of PNC. Compared to the traditional transmission and the straightforward network coding, our analytical results show that PNC can improve the network throughput by a factor of 2 and 1.5 respectively for the 1-D network, and by a factor of 3 and 2 respectively for the 2-D network.

*Related Work:*

In 2006, we proposed PNC in [6] as demodulation

This work was partially supported by the Competitive Earmarked Research Grant (Project Number 414507) established under the University Grant Committee of the Hong Kong Special Administrative Region, China.

Shengli Zhang is with the Information Engineering Department, the Chinese University of Hong Kong. (e-mail: slzhang@ ie.cuhk.edu.hk).

Soung Chang Liew is with the Information Engineering Department, the Chinese University of Hong Kong. (e-mail: soung@ ie.cuhk.edu.hk).

Part of this work was published in ACM Mobicom 2006.

mappings based on different modulation schemes. A similar idea was also published independently in [7] at the same time by another group. After that, a large body of work from other researchers on PNC began to appear. The work can be roughly divided into three categories.

In the first category, PNC is regarded as a modulation-demodulation technique. Many new PNC mapping schemes have been proposed since [6]. For example, [8] proposed a scheme based on Tomlinson-Harashima precoding. Following [6], ref. [9] proposed a simple relay strategy called analog network coding (ANC), in which the relay amplifies and forwards the received superimposed signal without any processing. Analog network coding turns out to be similar to a scheme earlier by researchers in the satellite communication society [10]. In [11], a number of memoryless relay functions, including PNC mapping and the BER optimal function, were identified and analyzed assuming phase synchronization between signals of the transmitters. In [12], we observed that there is a one-to-one correspondence between a relay function and a specific PNC scheme under the general definition of memoryless PNC. Besides the precise definition of memoryless PNC which distinguishes it from the traditional straightforward network coding (SNC), [12] also gave a number of new PNC schemes. Ref. [13] proposed a new PNC scheme where the relay maps a group constellation points to one signal according to the phase difference of the two end nodes' signals. The mechanism also takes care of the phase difference between the two end nodes implicitly.

In the second category, PNC and channel coding are studied jointly. In [14-16], PNC was combined with Lattice code or LDPC code. It was proved that the capacity of the two-way relay channel can be approached in high SNR and low SNR. In [14-16], channel coding and PNC mapping are performed independently (i.e., successively). In [17], we proposed a novel scheme which treats channel coding and PNC in an integrated manner. We show that joint channel-PNC decoding can outperform the previous schemes significantly.

In the third category, the focus is on the performance impact and significance of PNC in large scale wireless networks. For one-dimensional wireless networks, [18] showed that PNC can improve the capacity by a fixed factor, although it does not change the scaling law. For two-dimensional wireless networks, [19] showed that PNC can increase capacity by a factor of 2.5 for the rectangular networks and a factor 2 for the hexagonal networks. However, the result in [18] is obtained based on a rough scheduling scheme which is established traditional network coding rather than physical layer network coding (the special properties of PNC is ignored). Our paper also discusses the application of PNC in large scale wireless networks. It is different from [18] in that we provide the construction of an explicit PNC-scheduling algorithm (specially designed for PNC), upon which all our results are established. Compared with [19], we consider the many-to-many scenario with multiple sources and destinations, while [19] only considered the one-to-many scenario with one source.

The rest of this paper is organized as follows. Section II overviews the basic idea of PNC with a linear 3-node multi-hop network. Section III and Section IV investigate the application of PNC in the 1-D regular linear network and 2-D regular grid network, respectively. Section VI concludes the paper.

## II. ILLUSTRATING EXAMPLE: A THREE-NODE WIRELESS LINEAR NETWORK

Consider the three-node linear network in Fig. 1. $N_1$ (Node 1) and $N_3$ (Node 3) are nodes that exchange information, but they are out of each other's transmission range. $N_2$ (Node 2) is the relay node between them.

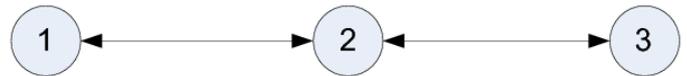

Figure 1.   A three-node linear network

This three-node wireless network is a basic unit for cooperative transmission and it has previously been investigated extensively [20-25]. In cooperative transmission, the relay node $N_2$ can choose different transmission strategies, such as Amplify-and-Forward or Decode-and-Forward [22], according to different Signal-to-Noise (SNR) situations. This paper focuses on the Decode-and-Forward strategy. We consider frame-based communication in which a time slot is defined as the time required for the transmission of one fixed-size frame. Each node is equipped with an omni-directional antenna, and the channel is half duplex so that transmission and reception at a particular node must occur in different time slots. Slow fading is assumed throughout this paper for the ease of synchronization.

Before introducing the PNC transmission scheme, we first describe the traditional transmission scheduling scheme and the "straightforward" network-coding scheme for mutual exchange of a frame in the three-node network [20, 25].

### A.    Traditional Transmission Scheduling Scheme

In traditional networks, interference is usually avoided by prohibiting the overlapping of signals from $N_1$ and $N_3$ to $N_2$ in the same time slot. A possible transmission schedule is given in Fig. 2. Let $S_i$ denote the frame initiated by $N_i$. $N_1$ first sends $S_1$ to $N_2$, and then $N_2$ relays $S_1$ to $N_3$. After that, $N_3$ sends $S_3$ in the reverse direction. A total of four time slots are needed for the exchange of two frames in opposite directions.

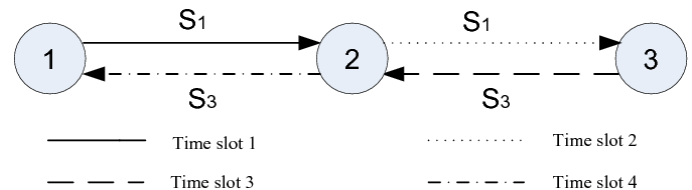

Figure 2.   Traditional scheduling scheme

## B. Straightforward Network Coding Scheme

Ref. [20] and [25] outline the straightforward way of applying network coding in the three-node wireless network. Fig. 3 illustrates the idea. First, $N_1$ sends $S_1$ to $N_2$ and then $N_3$ sends frame $S_3$ to $N_2$. After receiving $S_1$ and $S_3$, $N_2$ encodes frame $S_2$ as follows:

$$S_2 = S_1 \oplus S_3 \qquad (1)$$

where $\oplus$ denotes bitwise exclusive OR operation being applied over the entire frames of $S_1$ and $S_3$. $N_2$ then broadcasts $S_2$ to both $N_1$ and $N_3$. When $N_1$ receives $S_2$, it extracts $S_3$ from $S_2$ using the local information $S_1$, as follows

$$S_1 \oplus S_2 = S_1 \oplus (S_1 \oplus S_3) = S_3 \qquad (2)$$

Similarly, $N_2$ can extract $S_1$. A total of three time slots are needed, for a throughput improvement of 33% over the traditional transmission scheduling scheme.

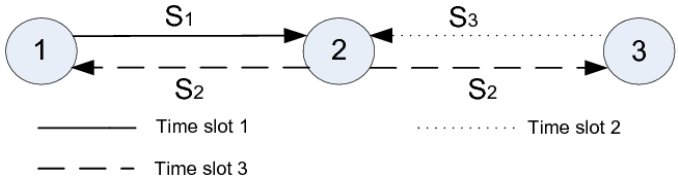

Figure 3.   Straightforward network coding scheme

## C. Physical-Layer Network Coding (PNC)

We now introduce PNC. Let us assume the use of BPSK modulation at all the nodes. We further assume symbol-level and carrier-phase synchronization, and the use of power control, so that the frames from $N_1$ and $N_3$ arrive at $N_2$ with the same phase and amplitude. The combined bandpass signal received by $N_2$ during one symbol period is

$$r_2(t) = s_1(t) + s_3(t) = a_1 \cos(\omega t) + a_3 \cos(\omega t)$$
$$= (a_1 + a_3)\cos(\omega t) \qquad (3)$$

where $s_i(t)$, $i = 1$ or 3, is the bandpass signal transmitted by $N_i$; $r_2(t)$ is the bandpass signal received by $N_2$ during one symbol period; $a_i$ is the BPSK modulated information bit of $N_i$; and $\omega$ is the carrier frequency. Then, $N_2$ will obtain a baseband signal $a_1 + a_3$.

Note that $N_2$ cannot extract the individual information transmitted by $N_1$ and $N_3$, i.e., $a_1$ and $a_3$, from the combined signal in $a_1 + a_3$. However, $N_2$ is just a relay node. As long as $N_2$ can transmit the necessary information to $N_1$ and $N_3$ for extraction of $a_1$, and $a_3$ over there, the end-to-end delivery of information will be successful. For this, all we need is a special modulation/demodulation mapping scheme, referred to as PNC mapping in this paper, to obtain the equivalence of GF(2) summation of bits from $N_1$ and $N_3$ at the physical layer.

*Table I* illustrates the idea of PNC mapping. In *Table I*, $s_j \in \{0, 1\}$ is a variable representing the data bit of $N_j$ and $a_j \in \{-1, 1\}$ is a variable representing the BPSK modulated bit of $s_j$ such that $a_j = 2s_j - 1$.

With reference to Table I, $N_2$ obtains the information bits:

$$s_2 = s_1 \oplus s_3 \qquad (4)$$

It then transmits

$$s_2(t) = a_2 \cos(\omega t) \qquad (5)$$

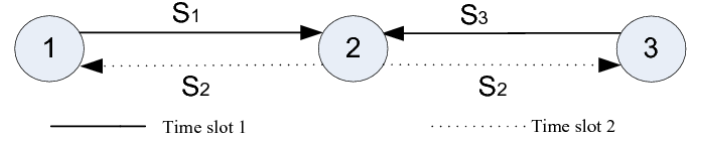

Figure 4.   Physical layer network coding

The BER analysis in [6] shows that the end-to-end BER for the three schemes is similar when the per-hop BER is low. Ignoring the slight BER difference, we have the following conclusion. For a frame exchange, PNC requires two time slots, 802.11 requires four, while straightforward network coding requires three. Therefore, PNC can improve the system throughput of the three-node wireless network by a factor of 100% and 50% relative to traditional transmission scheduling and straightforward network coding, respectively.

TABLE I. PNC MAPPING: MODULATION MAPPING AT $N_1$, $N_2$; DEMODULATION AND MODULATION MAPPINGS AT $N_3$

| Modulation mapping at $N_1$ and $N_3$, | | | | Demodulation mapping at $N_2$ | | | |
|---|---|---|---|---|---|---|---|
| | | | | Input | Output | | |
| Input | | Output | | | | Modulation mapping at $N_2$ | |
| | | | | | | Input | Output |
| $s_1$ | $s_3$ | $a_1$ | $a_3$ | $a_1 + a_3$ | $s_2$ | | $a_2$ |
| 1 | 1 | 1 | 1 | 2 | 0 | | -1 |
| 0 | 1 | -1 | 1 | 0 | 1 | | 1 |
| 1 | 0 | 1 | -1 | 0 | 1 | | 1 |
| 0 | 0 | -1 | -1 | -2 | 0 | | -1 |

## III.   APPLYING PNC IN REGULAR 1-D NETWORKS

Our discussions so far has only focused on the simple 3-node network with one bidirectional flow. In this section, we discuss the application of PNC in more general networks.

### A. Regular linear network with one bidirectional flow

Consider a regular linear network with $N$ nodes with equal spacing between adjacent nodes. Label the nodes as node 1, node 2, …, node $N$, successively with nodes 1 and $N$ being the two source and destination nodes, respectively. Fig. 5 shows a network with $N = 5$. Suppose that node 1 is to transmit frames $X_1$, $X_2$, …. to node $N$, and node $N$ is to transmit frames $Y_1$, $Y_2$, …. to node 1.

We could divide the time slots into two types: odd slots and even slots. In the odd time slots, the odd-numbered nodes transmit and the even-numbered nodes receive. In the even time slots, the even-numbered nodes transmit and the odd-numbered nodes receive.

Fig. 5 shows the sequence of frames being transmitted by the nodes in a 5-node network. In slot 1, node 1 transmits $X_1$ to node 2 and node 5 transmits $Y_1$ to node 4 at the same time. In slot 2, node 2 and node 4 transmit $X_1$ and $Y_1$ to node 3 simultaneously; both node 2 and node 4 also store a copy of $X_1$ and $Y_1$ in their buffer respectively. In slot 3, node 1 transmits $X_2$ to node 2, node 5 transmits $Y_2$ to node 4 and node 3 broadcasts $X_1 \oplus Y_1$ simultaneously; node 3 stores a copy of $X_1 \oplus Y_1$ in its buffer. Adding the stored $X_1$ to $X_2 \oplus X_1 \oplus Y_1$ received with PNC detection, node 2 can obtain $Y_1 \oplus X_2$. Node 4 can obtain $Y_2 \oplus X_1$ similarly. In slot 4, node 2 and node 4 broadcast $Y_1 \oplus X_2$ and $Y_2 \oplus X_1$ respectively. In this way, node 5 receives a copy of $X_1$ and node 1 receives $Y_1$ in slot 4. Also, in slot 4, node 3 obtains $Y_2 \oplus X_2$ by adding stored packet $X_1 \oplus Y_1$ to the received packet $X_1 \oplus Y_2 \oplus X_2 \oplus Y_1$.

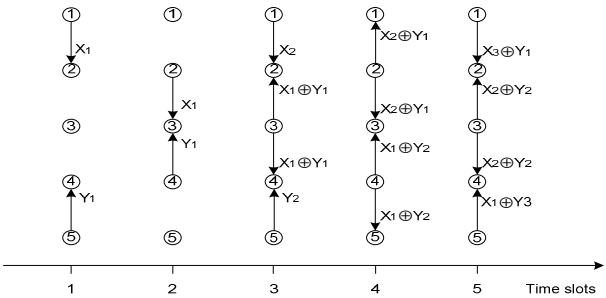

Figure 5.    Bidirection PNC transmission in linear network

With reference to Fig. 5, we see that a relay node forwards two frames, one in each direction, every two time slots. So, the throughput is 0.5 frame/time slot in each direction. Due to the half duplex assumption, this is the maximum possible throughput we can achieve.

As detailed above, when applying PNC on the linear network, each node transmits and receives alternately in successive time slots; and when a node transmits, its adjacent nodes receive, and vice versa (see Fig. 5). Let us investigate the signal-to-inference ratio (SIR) given this transmission pattern to make sure that it is not excessive. Consider the worst-case scenario of an infinite chain. We note the following characteristics of PNC from a receiving node's point of view:

a) The interfering nodes are symmetric on both sides.
b) The simultaneous signals received from the two adjacent nodes do not interfere due to the nature of PNC.
c) The nodes that are two hops away are also receiving at the same time, and therefore will not interfere with the node.

Therefore, the two nearest interfering nodes are three hops

away. We have the following SIR:

$$SIR = \frac{P_0 / d^\alpha}{2 * \sum_{l=1}^{\infty} P_0 / [(2l+1)d]^\alpha} \qquad (6)$$

where $P_0$ is the common transmitting power of nodes and $\alpha$ is the path-loss exponent. Assume the two-ray transmission model where $\alpha = 4$. The resulting SIR is about 16dB and the impact of the interference on BER is negligible for BPSK based on [26] (the capture threshold is often set to 10dB in wireless networks [3]). More generally, a thorough treatment should take into account the actual modulation scheme used, the difference between the effects of interference and noise, and whether or not channel coding is used. However, we can conclude that as far as the SIR is concerned, PNC is not worse than *traditional scheduling* (see Section V) when generalized to the $N$-node linear network.

### B.  Regular linear network with multiple flows

Part A considers only one bidirectional flow. Here we consider a general setting in which there are $K$ unidirectional flows in the $N$-node linear network. Note that this generalization includes the scenario in which there is a combination of unidirectional and bidirectional flows in the network, since each bidirectional flow can be considered as two unidirectional flows.

To allow PNC to be applied, we compose bidirectional flows out of the $K$ unidirectional flows by matching pairs of unidirectional flows in opposite directions. The bidirectional flows can then make use of PNC for transmission, while the remaining unmatched unidirectional flows make use of the traditional strategy of multi-hop data transmission.

The optimal way to compose the bidirectional flows and schedule the transmission of the links in the flows is a tough problem. Here we consider a simple heuristic which is asymptotically optimal for the regular $N$-node linear network when $N$ goes to infinity as shown in Part C. For simplicity, we assume all flows have equal traffic.

We define the following terms with respect to the linear network. Let us label the nodes from left to right by 1 to $N$ sequentially. Let $(s_i, d_i)$ denote the source-destination pair of flow $i$. For a right-bound flow, $s_i < d_i$; for a left-bound flow, $s_i > d_i$. Let $F$ denote the overall set of flows, and $F_R \subseteq F$ be the set of right-bound flows and $F_L \subseteq F$ be the set of left-found flows.

Two right-bound (left-bound) flows $i$ and $j$ are said to be *non-overlapping* if $d_i < s_j$ or $d_j < s_i$ ( $s_i < d_j$ or $s_j < d_i$ ). A *right packing* (*left packing*) is a set of non-overlapping right-bound flows (left-bound flows). A dual packing consists of a right packing and a left packing. Fig. 6 shows an example of a dual packing. Flows 2 and 3 form a right packing, and Flow 1 forms a left packing. Note that some of the nodes are traversed by both a right-bound flow and a left-bound flow. Let us call these nodes the common nodes, and the other nodes the non-common nodes. A sequence of adjacent common nodes, flanked by but not including two non-common nodes at two ends (an ellipse in Fig. 6), forms a *PNC unit*, and we can use the PNC mechanism for transporting the bidirectional traffic over it. A sequence of adjacent non-common nodes, together with the two common nodes flanking them (a rectangle in Fig. 6), may or may not have traffic flowing over them. When there is traffic, the traffic is in one direction only, and the traditional multi-hop communication technique can be used to carry the unidirectional traffic. Essentially, by forming a dual packing, we also form many "virtual" bidirectional flows (each corresponding to a PNC unit) on which PNC can be applied.

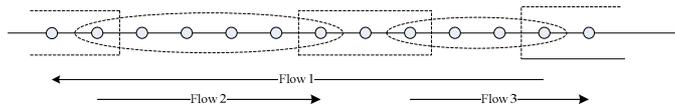

Figure 6.   An example of a dual packing formed by a right packing and a left packing. An ellipse corresponds to a PNC unit. The nodes between two adjacent ellipses (including the terminal nodes of the ellipses) are grouped together by a rectangle.

Our heuristic as follows consists of a method of forming dual packings from the $K$ unidirectional flows:

while ( $F \neq \varnothing$ ) { /* Each iteration in the while loop forms a dual packing. */

while ( $F_R \neq \varnothing$ ) { /* Each iteration in the while loop tries to find a "tight" right packing */

largestDest=0;

while (true) {

/* Each iteration in the while loop includes one more flow into the right packing being assembled. */

$$i = \arg \min_{j \in F_R : s_j > \text{largestDest}} s_j$$

/* Select a flow with the smallest source larger than LargestDest; assume "null" is returned if there is no more flow left in $F_R$ with $s_j > \text{largestDest}$ . */

if ( $i \neq$ null ) {

include flow $i$ into the current right packing being assembled;

largestDest $= d_i$ ;

remove flow $i$ from $F$;

} else

break;

/* Break out of the while(true) loop. */

}

}

while ( $F_L \neq \varnothing$ ) {

/* Each iteration in the while loop tries to find a "tight" left packing. */

/* Comment: details omitted here; the procedure is similar to the "while ( $F_R \neq \varnothing$ )" loop above except that largestDest is replaced by smallestDest; $s_j > \text{largestDest}$ is replaced by $s_j < \text{smallestDest}$ etc. */

}

/* Combine the right packings and left packings one by one to obtain dual packings */

}

The dual packings yield a set of "virtual" bidirectional flows, each corresponding to a PNC unit. Scheduling can then be performed as follows. Let us refer to the time needed for all the $K$ unidirectional flows to transfer one packet from source to destination as one *frame*. Each link (hop) of a flow is allocated one time slot for transmission within a frame. A frame is further divided into two intervals, as follows:

1) The first interval is dedicated to the PNC units (i.e., ellipses). Note that if there are $M$ dual packings, $2M$ time slots are needed in the worst case; in the worst case, different dual packings use different time slots to transmit, and 2 time slots are needed for each dual packing[1].

2) The second interval is dedicated to the non-PNC units (i.e., rectangles). The nodes of all rectangles of all dual packings are scheduled to transmit using the conventional scheme.

The number of time slots needed in the second interval depends on both the number and the lengths of the rectangles. As will be shown in Part C, it can be ignored compared to the time slots needed in the first interval as $N$ goes to infinity.

*C.  Throughput of 1-D network with PNC*

We now show that the packing and scheduling strategies presented in Part B can allow the upper-bound capacity of 1-D network to be approached when the number of nodes $N$ goes to infinity. Furthermore, compared with the conventional schemes discussed in [27], PNC can achieve a constant factor of throughput improvement.

We first detail the system model. To avoid edge effects, we consider a "large" circle instead of a line. The $N$ nodes are uniformly distributed over the circle with a constant distance between adjacent nodes. Without loss of generality, let the distance between two adjacent nodes be a unit distance. Each transmission is over only one unit distance (i.e., a node only transmits to its two adjacent nodes). Consider the receiver of a link. We assume that simultaneous transmission by another link whose transmitter is two or more hops away from the receiver of the first link will not cause a collision to the first link. In our model, $N/2$ nodes are randomly chosen as the source nodes. The remaining $N/2$ nodes are potential destination nodes. For each source node, a unique destination node is chosen among the $N/2$ potential destination nodes with equal probability. We assume matching without replacement in that the destination node chosen for a source node will not be put back to the pool before the destination node of another source is chosen. The route for a source-destination pair is also predetermined in a random way (note: there are two routes from a source to its destination, one in the clockwise direction and the other in the counterclockwise direction).

---

[1]  Two caveats are in order. The first is that according to our construction, there could be "trivial" PNC units with two nodes only. In this case, the PNC mechanism is not needed, and each node gets to transmit directly to the other node. Regardless of whether the PNC unit is trivial or not, two time slots are needed for the bidirectional flows. The second caveat is that there could be two PNC units in the same dual packing next to each other. For example, suppose nodes 1, 2, and 3 form a PNC unit, and nodes 4, 5, 6 forms another. To avoid conflict, the scheduling of the transmissions on these two PNC units should be such that nodes 1, 3, 4 and 6 transmit in one time slot while nodes 2 and 5 transmits in another time slot. Again, two time slots are needed.

The analytical results for the traditional transmission scheme and straightforward network coding scheme in our circular model are similar to those in the 1-D linear network in [27] when $N$ goes to infinity. Using similar approach, it is not difficult to obtain the respective per-flow throughputs in our circular network as

$$\lambda_T(N) = \frac{2}{N} \qquad \lambda_S(N) = \frac{8}{3N} \qquad (7)$$

where unit link bandwidth is assumed.

Let us now focus on the PNC throughput. We will show that PNC can achieve the per-flow throughput $4/N - \varepsilon$ for any small positive value $\varepsilon$ as $N$ goes to infinity. Let us first provide further details to the scheduling strategy presented in Part B.

The packing and scheduling are as follows. For packing, we first unwrap the circle to a non-circular linear network by randomly selecting the source node of a clockwise flow, labelled $s$, on the circle as the start point of the linear network. The adjacent node of the selected source node in the counterclockwise direction in the circle, labeled $e$, will serve as the end point of the linear network. Next, we obtain one packing of the clockwise flows according to the packing algorithm in Part B. It is possible that the last selected flow crosses the start point. In that case, we cut the flow into two sub-flows by performing the cut between the start point and the end point, and only consider the first sub-flow in the aforementioned packing. After forming the above clockwise unidirectional packing, we form a matching counterclockwise unidirectional packing at choosing $e$ as the start point and $s$ as the end point. If there is an existing counterclockwise flow with $e$ as its source node, we will start with this flow in the unidirectional packing. If not, we will choose the next flow with source node closest to $e$ in the counterclockwise direction in our packing.

For "traffic balance", after getting the first dual packing as above, for the next dual packing, we will start with forming the counterclockwise unidirectional packing first (i.e., $s$ and $e$ will be defined with respect to the counterclockwise packing) before constructing the matching clockwise packing. Repeating the above procedure allows us to form a series of dual packings.

The scheduling of transmissions is the same as that in Part B except that here we also have to consider the transmission across the two sub-flows cut as above, if any. We assume the traffic from the destination of a preceding sub-flow to the source of its corresponding sub-flow is transmitted using the conventional scheme in the second interval.

With the above packing and scheduling strategies, we have the following theorem on the per-flow throughput of the 1-D circular network when $N$ goes to infinity.

*Theorem 1:* With PNC, we can approach the upper bound of the per-flow throughput of the 1-D network:

$$\lambda_P(N) = \frac{4}{N} \qquad (8)$$

*Sketch of Proof:* A sketch of the proof for *Theorem 1* is provided here and a detailed proof is given in the Appendix. With the help of the max-flow min-cut theorem, the upper bound of the per-flow throughput for our 1-D circular network can be shown to be $4/N$. That this upper bound can be approached with the application of the aforementioned PNC packing and scheduling strategies is argued as follows. Consider the original $N/4$ unidirectional flows. With PNC packing and scheduling, these flows have been decomposed into PNC units and non-PNC units for transmission in the first and second intervals. For each round of first and second intervals (i.e., for each frame), one packet is transported from the source to the destination of each flow. We can show that the number of time slots needed in the first interval for all the flows is at most $(1+\varepsilon_1)N/4$, where the small positive quantity $\varepsilon_1$ goes to zero as $N$ goes to infinity. The number of time slots needed in the second interval, on the other hand, is $\varepsilon_2 N$, where the small positive quantity $\varepsilon_2$ goes to zero as $N$ goes to infinity. Then we can obtain the per-flow throughput with PNC: $1/\left(N/4 + \varepsilon_1 N/4 + \varepsilon_2 N/4\right) = (1-\varepsilon)N/4$.

A corollary of *Theorem 1* is that PNC can improve the throughput of the 1-D network by a factor of 2 and 1.5 relative to the traditional transmission scheme and the SNC scheme (7), respectively.

## IV. APPLYING PNC IN 2-D GRID NETWORK

Section III focused on the 1-D regular network. This section investigates the application of PNC in a 2-D regular gird network. We assume the same transmission protocol as in section III.

### A. 2-D Grid Network with one bidirectional flow in each line

Fig. 7 shows the grid network under consideration, in which $N$ nodes are uniformly located at the cross points as shown. In this part, we first consider the case in which each line (horizontal or vertical) on the grid has one and only one bidirectional flow. Specifically, the two end nodes in each line, node 1 and node $\sqrt{N}$ , exchange information through the relay nodes in between.

The flows transmit with the following PNC schedule. Consider the horizontal lines (similar schedule applies for the vertical lines). The first two time slots are dedicated to transmissions on lines $1, J+1, 2J+1,...$ ; the next two time slots are dedicated to transmissions on lines nodes on the lines $2, J+2, 2J+2,...$ ; and so on. The separation $J$ must be large enough for acceptable SIR. In the example of Fig. 7, $J$=4.

For a group of simultaneous active lines, to reduce SIR, when the odd nodes transmit on one active line, then the even nodes will transmit on its two adjacent active lines, as shown in Fig. 7.

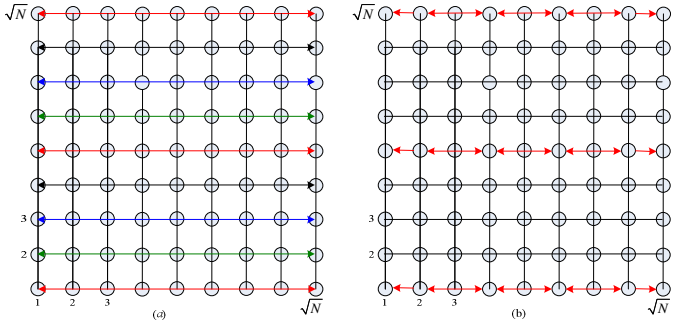

Figure 7. Subfigure (a) shows 2-D grid network with one bidirectional flow in each line. The lines separated by $J$−1=3 lines, i.e., the lines with the same color, are allowed to transmit simultaneously. Subfigure (b) shows a scheduling for one group active lines (red lines) in a specific time slot.

Let us investigate the SIR of this transmission pattern given a $J$. Consider the worst-case scenario in which $N$ goes to infinity. For a given receiver, the interference from the nodes within the same line is $I_1 = 2*\sum_{l=1}^{\infty} P_0/[(2l+1)d]^{\alpha}$ , where $P_0$, $l$, $d$=1, and $\alpha$ are defined similarly as in section III-A. Without loss of generality, suppose that the receiver is an even node. The interference from the other active lines whose odd nodes are transmitting is $I_2 = 4\sum_{k=0}^{\infty}\sum_{l=0}^{\infty}\dfrac{P_0}{[(2l)^2 d^2 + J^2(2k+2)^2 d^2]^{\alpha/2}}$ , and the interference from the other active lines whose even nodes are transmitting is $I_3 = 4\sum_{k=0}^{\infty}\sum_{l=0}^{\infty}\dfrac{P_0}{[(2l+1)^2 d^2 + J^2(2k+1)^2 d^2]^{\alpha/2}}$ . Thus, the overall SIR is given by

$$SIR = \frac{P_0/d^{\alpha}}{I_1 + I_2 + I_3} \qquad (9)$$

Assuming $\alpha = 4$ , the SIR in (9) is about 13.5dB, 12.3dB, and 10.0dB for $J$ equals 5, 4, and 3 respectively. With an assumed 10dB target, $J$=3 is enough to guarantee successful transmission.

## B. 2-D Grid network with multiple random flows

Let us now investigate the application of PNC in the 2-D grid network with a more general traffic pattern. With respect to Fig. 7, we now randomly choose $N/2$ of the nodes as the source nodes. The remaining $N/2$ nodes are the destination nodes.

Here we apply a simple routing scheme, as in [27]. For a source-destination pair at positions $(x_s, y_s)$ and $(x_d, y_d)$, the data will first be forwarded vertically to the node at $(x_s, y_d)$ before being forwarded horizontally to the destination. The horizontal and vertical transmissions are separated into two different time intervals. For horizontal (or vertical) transmissions, the scheduling within each line (column) is the same as that in the section III-B and the scheduling among different lines (columns) is the same as in part A.

When $N$ goes to infinity, the number of nodes in each line or column, $\sqrt{N}$ , also goes to infinity, and the per-flow PNC throughput in each line or column will approach $4/\sqrt{N}$ , as argued in section III. Since the horizontal transmission and vertical transmission are scheduled in different time interval and in each interval every $J$ lines (columns) transmit simultaneously, the per-flow transmission of PNC in the 2-D grid network can approach

$$\lambda_p(N) = \frac{4}{\sqrt{N}} \cdot \frac{1}{J} \cdot \frac{1}{2} = \frac{2}{J\sqrt{N}} \qquad (10)$$

For comparison purposes, let us look at the per-flow throughput under the traditional transmission strategy and under the straightforward network coding strategy. With the routing/scheduling strategy and the corresponding throughput analysis in [27], we can show that the traditional transmission scheme and SNC scheme can achieve the following throughputs, respectively:

$$\lambda_T(N) = \frac{4}{(1+\Delta)\sqrt{N}} \cdot \frac{1}{3} \cdot \frac{1}{2} = \frac{2}{9\sqrt{N}}$$

$$\lambda_C(N) = \frac{4}{(1+\Delta/2)\sqrt{N}} \cdot \frac{1}{3} \cdot \frac{1}{2} = \frac{1}{3\sqrt{N}} \qquad (11)$$

In the 2-D grid network, the nodes are tightly packed than in the 1-D network, and the interfering nodes must be kept at least 3 hops away, i.e., $\Delta = 2$ , to obtain an SIR of no less than 10dB (note: in the 1-D network, $\Delta$ could be 1 for SIR of about 10dB). When $\Delta = 2$ , we can verify throughputs better than (11) cannot be achieved. In other words, the throughput in (11) is also the upper bound for traditional transmission scheme and SNC scheme under all possible schedulings.

Therefore, setting $J$=3 in (10), we conclude that PNC can achieve a throughput improvement factor of 3 and 2 relative to the traditional transmission scheme and the SNC scheme, respectively. Note that the improvement factors under the 2-D network are larger than those under the 1-D network, which are 2 and 1.5, respectively (see section III).

## V. CONCLUSION

This paper has introduced a novel scheme called *Physical-layer Network Coding* (PNC) that significantly enhances the throughput performance of multi-hop wireless networks. Instead of avoiding interference caused by simultaneous electromagnetic waves transmitted from multiple sources, PNC embraces interference to effect network-coding operation directly from physical-layer signal modulation and demodulation. With PNC, signal scrambling due to interference, which causes packet collisions in the MAC layer protocol of traditional wireless networks (e.g., IEEE 802.11), can be eliminated.

We have proposed explicit scheduling algorithms for PNC in 1-D and 2-D regular networks with multiple random flows. It is shown that PNC can potentially achieve 100% and 50% throughput increases compared with traditional transmission and straightforward network coding, respectively, in the 1-D regular linear network. The throughput improvements are even larger in the 2-D regular network: 200% and 100%, respectively. In particular, PNC can allow the upper-bound throughput of the 1-D regular network to be approached as the number of nodes goes to infinity.

## Appendix: Proof of *Theorem 1*

This appendix proves *Theorem 1* in three steps. First, the fact that $4/N$ is the upper bound for the throughput of the 1-D circular linear network can be argued as follows. Let us consider the number of time slots needed so that each flow can transport one packet from its source to its destination. Due to half-duplexity, there can be at most $N/2$ transmitting nodes in a time slot. In general, each transmitting node can transmit to at most two of its adjacent nodes simultaneously. Hence, in total, there can be at most $N$ one-hop transmissions being successfully completed in each time slot. The number of hops between the source and destination of a flow is on average $N/2$. There are altogether $N/2$ flows. Using Chernoff bound, we can show that the total number of one-hop transmissions required (aggregated over all flows) is $N^2/4$ w.h.p. as $N$ goes to infinity. Thus, the time slots needed is lower bounded by $\dfrac{N^2/4}{N} = N/4$. Within this number of time slots, each flow transports a packet from source to destination. Thus, the per-flow throughput is upper bounded by $\lambda \le \dfrac{1}{N/4} = 4/N$.

Next, we prove that the number of time slots needed in the second interval is negligible compared to $N$, denoted by $\varepsilon_2 N$ where $\varepsilon_2$ is a small positive quantity that goes to zero as $N$ goes to infinity. The total one-hop transmissions in the second interval can be divided into two parts, the one-hop transmissions in the rectangles and the one-hop transmissions between sub-flows (created when we unwrap the circular network into a linear network).

Let us first consider the rectangles. As shown in Fig. A-1, within a dual packing, the rectangles do not overlap. Furthermore, the two end nodes in a rectangle must be either a source or destination node of some flow. As a proof technique, let us artificially divide the rectangles into two groups according to the dual packings containing them. Recall that the dual packings are formed successively in our packing algorithm. Consider the first $(1-\varepsilon_3)$ fraction of all flows (including the original flows and the generated sub-flows) that are included successively into the dual packings. The first group of rectangles arises from these flows. The second group of rectangles belongs to remaining $\varepsilon_3$ fraction of the flows. We set $\varepsilon_3$ such that $\varepsilon_3 = 1/\sqrt{\log N}$.

As discussed in Section III-B, when we perform packing on the circular network by unwrapping it to a linear network, it is possible for a flow to be cut into two subflows. Each clockwise unidirectional packing contains at least one flow that does not generate subflows (a flow cannot have more than $N$ hops). As a corollary, if the clockwise packing contains a flow that has been cut into two subflows, then the packing must contain at least two flows to start with. One of these subflows will be relegated to a future packing exercise. So, each clockwise packing reduces the number of remaining flows to be packed by at least one. For the matching counterclockwise packing, at most one flow will be cut into two subflows. Thus, the matching counterclockwise packing does not increase the number of remaining counter-clockwise flow. Recall from the discussion in Section III-B that for "traffic balance" successive dual packings will start with clockwise and counterclockwise packings in an alternate manner. Thus, successive dual packings will reduce the numbers of remaining clockwise and counterclockwise flows by at least one alternately.

In the beginning, there are $N/2$ original flows ($N/4$ of which are clockwise and $N/4$ of which are counterclockwise flows). From the argument in the previous paragraph, there are altogether at most $N/2$ dual packings. Each dual packing will at most generate at most two extra flows to the flow pool (because of cut between $s$ and $e$). Thus, altogether there could be at most $N$ extra flows being generated. Hence, the total number of flows (including the original flows and the subflows) is $3N/2$.

In general, since the two end nodes of a rectangle must be either a source or a destination of some flow, the number of rectangles in a dual packing is no more than the number of flows in that dual packing (note: some non-end nodes within a rectangle could also be sources or destinations; thus the "no more than" rather than "equal to"). Therefore, the number of rectangles in the first group is therefore no more than $(1-\varepsilon_3)N$. For these rectangles, as shown in *Lemma 2* at the end of this appendix, the number of nodes in each group-1 rectangle is no more than $(1-\varepsilon_4)\log(N) + \varepsilon_4 N$ w.h.p., where $\varepsilon_4$ is a small positive quantity that goes to zero when $N$ goes to infinity. Similarly, the number of rectangles in the second group is upper bounded by $\varepsilon_3 N$. As a trivial bound, we will upper-bound the number of nodes in each group-2 rectangle by $N$. Note that each node will at most transmit once within a rectangle (group-1 or group-2) for traffic forwarding. Thus, the total number of one-hop transmissions needed for the rectangles is upper bounded by

$$T_1 = (1-\varepsilon_3)N \cdot \left[(1-\varepsilon_4)\log(N) + \varepsilon_4 N\right] + \varepsilon_3 N \cdot N .$$
(A-1 )

Now, consider the transmissions across sub-flows. A one-hop transmission is needed for two adjacent sub-flows generated by the cut when we unwrap the circular network to a corresponding linear network. In other words, there is a one-hop transmission whenever there is an extra sub-flow, which is upper bounded by $N/2$ according to the above argument. Thus, the total number of one-hop transmissions between all adjacent sub-flows is upper bounded by $T_2 = N/2$.

Putting things together, the total one-hop transmissions in the second interval is upper bounded by $T_1 + T_2$. Since we determine the start and end nodes of each dual packing in a uniformly random way and pack each unidirectional packing in a uniformly random way, the one hop transmissions in the rectangles are also uniformly distributed among all the $N$ nodes along the circle. With the traditional transmission scheme, there are $N/2$ one-hop transmissions in each time slot. Therefore, the time slots needed in the second interval is upper bounded by

$$
\begin{aligned}
k_2 &= \frac{T_1 + T_2}{N/2} \\
&= \frac{(1-\varepsilon_3)N \cdot \left[(1-\varepsilon_4)\log(N) + \varepsilon_4 N\right] + \varepsilon_3 N \cdot N + N/2}{N/2} \\
&= 2(1-\varepsilon_3)(1-\varepsilon_4)\log(N) + 2(1-\varepsilon_3)\varepsilon_4 N + \varepsilon_3 N + 1 \\
&= N\varepsilon_2
\end{aligned}
$$
(A-2)

where $\varepsilon_2$ is determined by $\varepsilon_3, \varepsilon_4$ and $N$. It is easy to show that $\varepsilon_2$ will go to zero as $N$ goes to infinity.

Finally, we prove that the number of time slots needed in the first interval is less than $(1+\varepsilon_1)N/4$. In a unidirectional packing, a residual node is an idle node that through which no packet passes (i.e., none of the flows of the unidirectional packing passes through the node). Thus, the number of nodes through which one packet passes in one unidirectional packing is $N$, minus the number of residual nodes. Consider a dual packing to which group-1 rectangles belong. According to *Lemma 1* immediately after the proof of Theorem 1 here, the number of residual nodes in each of the unidirectional packings of the dual packings is less than $\log(N)$ *w.h.p.*. That is, the number of non-residual nodes in a unidirectional packing is more than $N$-$\log(N)$ w.h.p., and the number of non-residual nodes in both the unidirectional packing of the dual packing is more than $2(N-\log N)$. That is, the traffic handled by each dual packing (in terms of packet flows across all nodes in the dual packing) is more than $2(N-\log N)$.

Now, consider an arbitrary node in the network. According to our model, it is either the source or destination of some flow. The packet of that flow passes through it with probability 1. For the other $N/2$ -1 original flows, a packet passes through the node with probability 1/2. By the Chernoff-Hoeffding theorem, the number of packets that go through each node is $\frac{1}{2} \cdot (N/2-1) + 1$ *w.h.p.*. Considering all $N$ nodes, the number of packets passing through them is $\left(\frac{1}{2}(N/2-1)+1\right)N$. Note that this is the total traffic which is more than the traffic in the dual packings to which group-1 rectangles belong.

Therefore, the number of dual packings to which the group-1 rectangles belong is upper bounded by

$$\left(\frac{1}{2}(N/2-1)+1\right)N \Big/ (2(N-\log(N))) \ w.h.p.. \quad \text{(A-3)}$$

Similar to the argument for group-1 rectangles, for the flows containing the group-2 rectangles, there are at most $\varepsilon_3 N$ flows which will generate at most $\varepsilon_3 N$ unidirectional packings, i.e., $\varepsilon_3 N/2$ dual packings. Then we can obtain that the total number of dual packings is no more than

$$\left(\frac{1}{2}(N/2-1)+1\right)N \Big/ (2(N-\log(N))) + \varepsilon_3 N/2 = (1+\varepsilon_1)N/8$$
(A-4)

with high probability, where $\varepsilon_1$ is determined by $\varepsilon_3$ and $N$. It is easy to verify that $\varepsilon_1$ goes to zero as $N$ goes to infinity. Since each packing needs at most two times slots, the time slots needed for the first interval is at most $k_1 = (1+\varepsilon_1)N/4$.

With the help of $k_1$ and $k_2$, we can obtain the lower bound of the per-flow throughput as

$$
\begin{aligned}
\lambda_P(N) &= \frac{1}{k_1 + k_2} = \frac{1}{(1+\varepsilon_1)N/4 + 2\log(N) + 2N\varepsilon_2 + 1} \\
&= \frac{4}{N}\frac{1}{1+\varepsilon_1 + 2\log(N)/N + 2\varepsilon_2 + 1/N} = \frac{4}{N}(1-\varepsilon)
\end{aligned}
$$
(A-5)

where $\varepsilon$ can be obtained from $\varepsilon_1, \varepsilon_2$ and $N$, and it goes to zero as $N$ goes to infinity. Then *Theorem 1* is proved.

***Lemma 1***: For any clockwise (counterclockwise) unidirectional packing contained in the dual packings to which group-1 rectangles belong, the number of residual nodes is less than $\log(N)$ w.h.p.

*Proof:* Let $P$ denote the set of dual packings to which group-1 rectangles belong. Let us focus on one clockwise unidirectional packing $p$ in $P$. The proof for the counterclockwise case is similar. Let $P_c$ be the clockwise packings in $P$. Let $m$ denote the number of clockwise flows in $P_c$. According to our way of partitioning the rectangles into the two groups, we have $m \le (1-\varepsilon_3)N_1$, where $N_1$ is the total number of clockwise flows.

Recall that in our traffic model, we randomly select $N/2$ nodes to be sources and $N/2$ nodes to be destinations. In other words, any node among the $N$ nodes is either a source or a destination. This applies to any residual node in $p$ as well. In particular, a residual node in $p$ is either 1) a destination node (of a clockwise or counter-clockwise flow); 2) a source node of a counter-clockwise flow; or 3) a source node of a clockwise flow. In case 3, since the residual node is a residual node in $p$, it must be a source node of a clockwise flow already packed (i.e., already belong to $P_c$) prior to packing $p$.

For a unidirectional packing, consider the first flow from the start point $s$. Suppose this flow ends at node $i$. Let us consider the probability of node $(i+1)$ being a residual node with respect to this unidirectional packing. Due to the randomness of our packing procedure and our random selection of sources and destinations for flows, node $(i+1)$ is a destination node with probability $p_1=1/2$, it is a source node of a counter-clockwise flow with probability $p_2=1/4$ w.h.p, and it is a source node of a pre-packed clockwise flow with probability $p_3 \le (1-\varepsilon_3)/4$ w.h.p. Then the probability that node $(i+1)$ is a residual node given that node $i$ is not a residual node is

$$P(1\,|\,0) = p_1 + p_2 + p_3 \le 1 - \varepsilon_3/4 \qquad \text{(A-6)}$$

In out notation above, the 1 in $P(1\,|\,0)$ refers to the fact that we have found one residual thus far, and the 0 refers to the fact that we have not found any residual node so far. Given node $(i+1)$ is a residual node, the probability that the node $(i+2)$ is also a residual node is $P(2\,|\,1) \le P(1\,|\,0)$ (due to sampling without replacement). The probability of a sequence of $l$ or more residual nodes is given by

$$P(1\,|\,0)P(2\,|\,1)P(3\,|\,2)\cdots P(l\,|\,l-1) \le \left[P(1\,|\,0)\right]^l \le \left[1-\varepsilon_3/4\right]^l \qquad \text{(A-7)}$$

When $l = \log(N)$, as $N$ goes to infinity, the above probability is $\exp(-\sqrt{\log(N)}/4)$, which will approach zero. Thus, *Lemma 1* is proved.

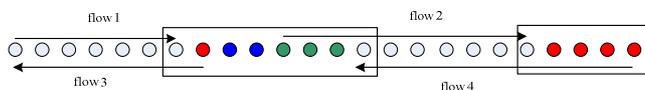

Figure A-1, An example of a dual packing, where flow 1 and flow 2 belong to the clockwise unidirectional packing, flow 3 and flow 4 belong to the counterclockwise unidirectional packing. The white nodes are non-residual nodes, the red nodes are the residual nodes of the clockwise unidirectional packing, the green nodes are the residual nodes of the counterclockwise packing and the blue nodes are the residual nodes of both the two unidirectional packings. The nodes in the rectangles are the uncommon nodes.

**Lemma 2:** For group-1 rectangles, the number of nodes in each rectangle is no more than $2\log(N)$ with probability $1-\varepsilon_4$, where $\varepsilon_4$ is a small positive quantity that goes to zero when $N$ goes to infinity.

*Proof:* With respect to Fig. A-1 and the explanation in its caption, let $N_r, N_g, N_b$ denote the number of red, green, and blue nodes in a dual packing, respectively. By Lemma 1, $N_r + N_b \le \log(N)$, and $N_g + N_b \le \log(N)$ w.h.p. Thus, $N_r + N_g + N_b \le N_r + N_g + 2N_b \le 2\log(N)$.